\documentclass[11pt]{article}
\usepackage{amsfonts}
\usepackage{amssymb}

\csname @addtoreset\endcsname{equation}{section}

\begin{document}
\begin{titlepage}
\begin{flushright}
CERN-TH/2000-187\\ LAPTH-803/2000 \\ hep-th/0007058
\end{flushright}
\vspace{.5cm}
\begin{center}
{\Large\bf Conformal superfields and BPS states in $AdS_{4/7}$
geometries \footnote[1]{To appear in the proceedings of the
Euroconference ``Noncommutative geometry and Hopf algebras in field
theory and particle physics", Torino, Villa Gualino (Italy),
September 20-30, 1999}}\\
 \vfill {\Large  Sergio Ferrara$^\dagger$ and Emery
Sokatchev$^\ddagger$ }\\
 \vfill  \vspace{6pt} $^\dagger$ CERN
Theoretical Division, CH 1211 Geneva 23, Switzerland
\\ \vspace{6pt}
$^\ddagger$ Laboratoire d'Annecy-le-Vieux de Physique
Th\'{e}orique\footnote[2]{UMR 5108 associ\'{e}e \`{a}
 l'Universit\'{e} de Savoie} LAPTH, Chemin
de Bellevue - BP 110 - F-74941 Annecy-le-Vieux Cedex, France

\end{center}
\vfill

\begin{center}
{\bf Abstract}
\end{center}
{\small We carry out a general analysis of the representations of
the superconformal algebras $\mbox{OSp}(8/4,\mathbb{R})$ and
$\mbox{OSp}(8^*/2N)$ in terms of harmonic superspace. We present a
construction of their highest-weight UIR's by multiplication of
the different types of massless conformal superfields
(``supersingletons").
\\ Particular attention is paid to the so-called ``short multiplets".
Representations undergoing shortening have ``protected dimension"
and may correspond to BPS states in the dual supergravity theory
in anti-de Sitter space.\\ These results are relevant for the
classification of multitrace operators in boundary conformally
invariant theories as well as for the classification of AdS black
holes preserving different fractions of supersymmetry.}
\end{titlepage}

\section{Introduction}

Superconformal algebras and their representations play a crucial
r\^{o}le in the AdS/CFT correspondence because of their dual r\^{o}le of
describing the gauge symmetries of the AdS bulk supergravity
theory and the global symmetries of the boundary conformal field
theory \cite{mal,gkp,wit}.

A special class of configurations which are particularly relevant
are the so-called BPS states, i.e. dynamical objects corresponding
to representations which undergo ``shortening".

These representations can only occur when the conformal dimension
of a (super)primary operator is ``quantized" in terms of the R
symmetry quantum numbers and they are at the basis of the
so-called ``non-renormalization" theorems of supersymmetric
quantum theories \cite{FIZ}.

There exist different methods of constructing the UIR's of
superconformal algebras. One is the so-called oscillator
construction of the Hilbert space in which a given UIR acts
\cite{bgg}. Another one, more appropriate to describe field
theories, is the realization of such representations on
superfields defined in superspaces \cite{SS,fwz}. The latter are
``supermanifolds" which can be regarded as the quotient of the
conformal supergroup by some of its subgroups.

In the case of ordinary superspace the subgroup in question is the
supergroup obtained by exponentiating a non-semisimple
superalgebra which is the semidirect product of a super-Poincar\'{e}
graded Lie algebra with dilatation ($\mbox{SO}(1,1)$) and the R
symmetry algebra. This is the superspace appropriate for non-BPS
states. Such states correspond to bulk massive states which can
have ``continuous spectrum" of the AdS mass (or, equivalently, of
the conformal dimension of the primary fields).

BPS states are naturally associated to superspaces with lower
number of ``odd" coordinates and, in most cases, with some
internal coordinates of a coset space $G/H$. Here $G$ is the R
symmetry group of the superconformal algebra, i.e. the subalgebra
of the even part which commutes with the conformal algebra of
space-time and $H$ is some subgroup of $G$ having the same rank as
$G$.

Such superspaces are called ``harmonic" \cite{GIK1} and they are
characterized by having a subset of the initial odd coordinates
$\theta$. The complementary number of odd variables determines the
fraction of supersymmetry preserved by the BPS state. If a BPS
state preserves $K$ supersymmetries then the $\theta$'s of the
associated harmonic superspace will transform under some UIR of
$H_K$.

For 1/2 BPS states, i.e. states with maximal supersymmetry, the
superspace involves the minimal number of odd coordinates (half of
the original one) and $H_K$ is then a maximal subgroup of $G$. On
the other hand, for states with the minimal fraction of
supersymmetry $H_K$ reduces to the ``maximal torus" whose Lie
algebra is the Cartan subalgebra of $G$.

It is the aim of the present paper to give a comprehensive
treatment of BPS states related to ``short representations" of
superconformal algebras for the cases which are most relevant in
the context of the AdS/CFT correspondence, i.e. the $d=3$ ($N=8$)
and $d=6$ (for arbitrary $N$). The underlying conformal field
theories correspond to world-volume theories of $N_c$ copies of
$M_2$ and $M_5$ branes in the large $N_c$ limit
\cite{AOY}-\cite{ckvp} which are ``dual" to AdS supergravities
describing the horizon geometry of the branes \cite{AGMOO}.

The present contribution summarizes the results which have
already appeared elsewhere \cite{FS2,FS3,FS4}. We first carry out
an abstract analysis of the conditions for Grassmann
(G-)analyticity \cite{GIO} (the generalization of the familiar
concept of chirality \cite{fwz}) in a superconformal context. We
find the constraints on the conformal dimension and R symmetry
quantum numbers of a superfield following from the requirement
that it do not depend on one or more Grassmann variables.
Introducing G-analyticity in a traditional superspace cannot be
done without breaking the R symmetry. The latter can be restored
by extending the superspace by harmonic variables
\cite{Rosly},\cite{GIK1},\cite{GIK11}-\cite{hh} parametrizing the
coset $G/H_K$. We also consider the massless UIR's
(``supersingleton" multiplets) \cite{ff2,Fr}, first as
constrained superfields in ordinary superspace
\cite{HSiT,bsvp,Howe} and then, for a part of them, as G-analytic
harmonic superfields \cite{GIK1,hh,HStT,Zp,Howe}. Next we use
supersingleton multiplication to construct UIR's of
$\mbox{OSp}(8^*/2N)$ and $\mbox{OSp}(8/4,\mathbb{R})$. We show
that in this way one can reproduce the complete classification of
UIR's of ref. \cite{Minw2}. We also discuss different kinds of
shortening which certain superfields (not of the BPS type) may
undergo. We conclude the paper by listing the various BPS states
in the physically relevant cases of $M_2$ and $M_5$ branes
horizon geometry where only one type of supersingletons appears.

Massive towers corresponding to 1/2 BPS states are the K-K modes
coming from compactification of M-theory on $AdS_{7/4}\times
S_{4/7}$ \cite{masstow,AOY}. Short representations of
superconformal algebras also play a special r\^ole in determining
$N$-point functions from OPE \cite{dHP,corbast}.

Another area of interest is the classification of AdS black holes
\cite{hawk}-\cite{duffl}, according to the fraction of
supersymmetry preserved by the black hole background.

In a parallel analysis with black holes in asymptotically flat
background \cite{FMG}, the AdS/CFT correspondence predicts that
such BPS states should be dual to superconformal states undergoing
``shortening" of the type discussed here.

\setcounter{equation}0
\section{The six-dimensional case}

In this section we describe highest-weight UIR's of the
superconformal algebras $\mbox{OSp}(8^*/2N)$ in six dimensions.
Although the physical applications refer to $N=1$ and $N=2$, it
is worthwhile to carry out the analysis for general $N$, along
the same lines as in the four-dimensional case \cite{AFSZ,FS1}. We
first examine the consequences of G-analyticity and conformal
supersymmetry and find out the relation to BPS states. Then we
will construct UIR's of $\mbox{OSp}(8^*/2N)$ by multiplying
supersingletons. The results exactly match the general
classification of UIR's of $\mbox{OSp}(8^*/2N)$ of Ref.
\cite{Minw2}.

\subsection{The conformal superalgebra $\mbox{OSp}(8^*/2N)$
and Grassmann analyticity}

The part of the conformal superalgebra $\mbox{OSp}(8^*/2N)$
relevant to our discussion is given below:
\begin{eqnarray}
  &&\{Q^i_\alpha, Q^j_\beta\} = 2\Omega^{ij}\gamma^\mu_{\alpha\beta}P_\mu
\;, \label{6.1}\\
  &&\{S^{\alpha\; i}, S^{\beta\; j}\} = 2\Omega^{ij}\gamma_\mu^{\alpha\beta}K^\mu
\;, \label{6.1'}\\
  &&\{Q^i_\alpha, S^{\beta\; j}\} = i\Omega^{ij}(\gamma^{\mu\nu})_\alpha{}^\beta
M_{\mu\nu} + 2\delta^\beta_\alpha (4T^{ij}-i\Omega^{ij}D)\;,
\label{6.2}\\
  &&[D,Q^i_\alpha]= {i\over 2}Q^i_\alpha\;, \qquad [D,S^{\alpha\; i}]= -{i\over 2}S^{\alpha\; i}\;, \label{6.2'}\\
  && [T^{ij},Q^k_\alpha]= -{1\over 2}(\Omega^{ki}Q^{j}_\alpha +
\Omega^{kj}Q^{i}_\alpha)\;, \label{6.3}\\
  && [T^{ij},T^{kl}]= {1\over 2}(\Omega^{ik}T^{lj} + \Omega^{il}T^{kj} +
\Omega^{jk}T^{li} + \Omega^{jl}T^{ki}) \;.\label{6.4}
\end{eqnarray}
Here $Q^i_\alpha$ are the generators of Poincar\'{e} supersymmetry
carrying a right-handed chiral spinor index $\alpha=1,2,3,4$ of
the Lorentz group $\mbox{SU}^*(4)\sim \mbox{SO}(5,1)$ (generators
$M_{\mu\nu}$) and an index $i=1,2,\ldots,2N$ of the fundamental
representation of the R symmetry group $\mbox{USp}(2N)$
(generators $T^{ij}=T^{ji}$); $S^{\beta\; j}$ are the generators
of conformal supersymmetry carrying a left-handed chiral spinor
index; $D$  is the generator of dilations, $P_\mu$ of
translations and $K_\mu$ of conformal boosts. It is convenient to
make the non-standard choice of the symplectic matrix
$\Omega^{ij}=-\Omega^{ji}$ with non-vanishing entries
$\Omega^{1\; 2N}=\Omega^{2\; 2N-1}=\ldots =\Omega^{N\; N+1}=1$.
The chiral spinors satisfy a pseudo-reality condition of the type
$\overline{Q_\alpha^i} = \Omega^{ij}Q_j^\beta c_{\beta\alpha}$
where $c$ is a $4\times 4$ unitary ``charge conjugation" matrix.
Note that the generators $M,P,K,D$ form the Lie algebra of
$\mbox{SO}(8^*)\sim \mbox{SO}(2,6)$ and the generators $Q,S$ form
an $\mbox{SO}(8^*)$ chiral spinor.

The standard realization of this superalgebra makes use of a
superspace with even coordinates $x^\mu$ and left-handed spinor
odd ones $\theta^{\alpha\; i}$. Unlike the four-dimensional case,
here chirality is not an option but is already built in. The only
way to obtain smaller superspaces is through Grassmann
analyticity. We begin by imposing a single condition of
G-analyticity:
\begin{equation}\label{6.6}
  q^1_\alpha\Phi(x,\theta)=0
\end{equation}
(here and in what follows the lower-case notation refers to the
matrix part of the generators). This condition amounts to
removing the odd variable $\theta^{\alpha\; 2N}$, i.e. to a
superspace with coordinates $x^\mu,\theta^{\alpha\;
1,2,\ldots,2N-1}$. From the definition of a superconformal
primary field we have
\begin{equation}\label{6.6'}
  s^{i\;\beta}\Phi=0\;,
\end{equation}
which, together with (\ref{6.6}) and the algebra
(\ref{6.1})-(\ref{6.4}) yields the consistency conditions
\begin{eqnarray}
  &&m_{\mu\nu}=0\;, \label{6.8}\\
  &&t^{11}=t^{12}=\ldots=t^{1\; 2N-1}=0\;, \label{6.9}\\
  &&4t^{1\; 2N}+\ell=0 \label{6.10}
\end{eqnarray}
(in (\ref{6.10}) $\ell$ denotes the conformal dimension, i.e. the
eigenvalue of $-iD$). Eq. (\ref{6.8}) implies that the superfield
$\Phi$ must be a Lorentz scalar. In order to interpret eqs.
(\ref{6.9}), (\ref{6.10}), we need the Cartan decomposition of
the algebra of $\mbox{USp}(2N)$ into:\\
 (i) raising operators
(corresponding to the positive roots):
\begin{equation}\label{6.111}
  T^{k\; 2N-l}\;, \ \ k=1,\ldots, N,\ l=k,\ldots,2N-k\quad (\mbox{simple if
$l=k$})\;;
\end{equation}
(ii) $[\mbox{U}(1)]^N$ charges:
\begin{equation}\label{6.112}
  H_k = -2 T^{k\; 2N-k+1}\;, \quad k=1,\ldots, N\;;
\end{equation}
(iii) the rest are lowering operators (corresponding to the
negative roots). The  Dynkin labels $a_k$ of a $\mbox{USp}(2N)$
irrep are defined as follows:
\begin{equation}\label{6.113}
  a_k=H_k-H_{k+1}\;, \ \ k=1,\ldots,N-1\;, \quad a_N=H_N\;,
\end{equation}
so that, for instance, the projection $Q^1$ of the supersymmetry
generator is the HWS of the fundamental irrep $(1,0,\ldots,0)$.

Now it becomes clear that (\ref{6.9}) is part of the
$\mbox{USp}(2N)$ irreducibility conditions whereas (\ref{6.10})
relates the conformal dimension to the sum of the Dynkin labels:
\begin{equation}\label{6.12}
  \ell = 2\sum_{k=1}^N a_k\;.
\end{equation}
Let us denote the highest-weight UIR's of the $\mbox{OSp}(8^*/2N)$
algebra by
$$
{\cal D}(\ell;J_1,J_2,J_3;a_1,\ldots,a_N)
$$
where $\ell$ is the conformal dimension, $J_1,J_2,J_3$ are the
$\mbox{SU}^*(4)$ Dynkin labels and $a_k$ are the $\mbox{USp}(2N)$
Dynkin labels of the first component. Then the G-analytic
superfields defined above are of the type
\begin{equation}\label{6.13}
 \Phi(\theta^{1,2,\ldots,2N-1}) \ \Leftrightarrow \
{\cal D}(2\sum_{k=1}^N a_k;0,0,0;a_1,\ldots,a_N)\;.
\end{equation}

The next step is to impose a second condition of G-analyticity
with the generator $q^2_\alpha$ which removes one more odd
variable and leads to a superspace with coordinates
$x^\mu,\theta^{\alpha\; 1,2,\ldots,2N-2}$. This implies the new
constraints
\begin{equation}\label{6.15}
  4t^{2\; 2N-1}+\ell=0\quad \Rightarrow \  a_1=0\;, \qquad t^{2\; 2N}=0\;.
\end{equation}
Note that the vanishing of the lowering operator $t^{2\; 2N}$
means that the subalgebra $\mbox{SU}(2)\subset \mbox{USp}(2N)$
formed by $t^{1\; 2N-1}$, $t^{2\; 2N}$ and $t^{1\; 2N}-t^{2\;
2N-1}$ acts trivially on the particular $\mbox{USp}(2N)$ irreps.
This is equivalent to setting $a_1=0$, as in (\ref{6.15}). Thus,
the new G-analytic superfields are of the type
\begin{equation}\label{6.17}
 \Phi(\theta^{1,2,\ldots,2N-2}) \ \Leftrightarrow \
{\cal D}(2\sum_{k=2}^N a_k;0,0,0;0,a_2,\ldots,a_N)\;.
\end{equation}

From (\ref{6.1}) it is clear that we can go on in the same manner
until we remove half of the $\theta$'s, namely
$\theta^{N+1},\ldots,\theta^{2N}$. Each time we have to set a new
Dynkin label to zero. We can summarize by saying that the
superconformal algebra $\mbox{OSp}(8^*/2N)$ admits the following
short UIR's corresponding to BPS states:
\begin{equation}\label{6.18}
  {p\over 2N}\mbox{ BPS}:\
{\cal D}(2\sum_{k=p}^N a_k;0,0,0;0,\ldots,0,a_p,\ldots,a_N)\;,
\quad p=1,\ldots,N\;.
\end{equation}

\subsection{Supersingletons}

There exist three types of massless multiplets in six dimensions
corresponding to ultrashort UIR's (supersingletons) of
$\mbox{OSp}(8^*/2N)$ (see, e.g., \cite{GT} for the case $N=2$).
All of them can be formulated in terms of constrained superfields
as follows.

{\sl (i)} The first type is described by a superfield
$W^{\{i_1\ldots i_n\}}(x,\theta)$, $1\leq n \leq N$, which is
antisymmetric and traceless in the external $\mbox{USp}(2N)$
indices (for even $n$ one can impose a reality condition). It
satisfies the constraint (see \cite{HSiT} and \cite{Park})
\begin{equation}\label{6.19}
  D^{(k}_\alpha W^{\{i_1)i_2\ldots i_n\}}=0 \qquad \Rightarrow \ {\cal
D}(2;0,0,0;0,\ldots,0,a_{n}=1,0,\ldots,0)
\end{equation}
where the spinor covariant derivatives obey the supersymmetry
algebra
\begin{equation}\label{6.21}
  \{ D^i_\alpha, D^j_\beta\} =
-2i\Omega^{ij}\gamma^\mu_{\alpha\beta}\partial_\mu\;.
\end{equation}
The components of this superfield are massless fields. In the case
$N=n=1$ this is the on-shell $(1,0)$ hypermultiplet and for
$N=n=2$ it is the on-shell $(2,0)$ tensor multiplet
\cite{HSiT,bsvp}.

{\sl (ii)} The second type is described by a (real) superfield
without external indices, $w(x,\theta)$. The corresponding
constraint is second-order in the spinor derivatives:
\begin{equation}\label{6.23}
 D^{(i}_{[\alpha} D^{j)}_{\beta]} w = 0 \qquad \Rightarrow \ {\cal
D}(2;0,0,0;0,\ldots,0)\;.
\end{equation}

{\sl (iii)} Finally, there exists an infinite series of
multiplets described by superfields with $n$ totally symmetrized
external Lorentz spinor indices,
$w_{(\alpha_1\ldots\alpha_n)}(x,\theta)$ (they can be made real
in the case of even $n$). Now the constraint takes the form
\begin{equation}\label{6.24}
  D^i_{[\beta}w_{(\alpha_1]\ldots\alpha_n)} = 0 \qquad \Rightarrow \ {\cal
D}(2+n/2;n,0,0;0,\ldots,0)\;.
\end{equation}

As shown in ref. \cite{FS3}, the six-dimensional massless
conformal fields only carry reps $(J_1,0)$ of the little group
$\mbox{SU}(2)\times \mbox{SU}(2)$ of a light-like particle
momentum. This result is related to the analysis of conformal
fields in $d$ dimensions \cite{Siegel1,AL}. This fact implies that
massless superconformal multiplets are classified by a single
$\mbox{SU}(2)$ and $\mbox{USp}(2N)$ R-symmetry and are therefore
identical to massless super-Poincar\'e multiplets in five
dimensions. Some physical implication of the above circumstance
have recently been discussed in ref. \cite{HULL2} where it was
suggested that certain strongly coupled $d=5$ theories effectively
become six-dimensional.

\subsection{Harmonic superspace}

The massless multiplets {\sl (i), (ii)} admit an alternative
formulation in harmonic superspace (see \cite{HStT,Zp,Howe} for
$N=1,2$). The advantage of this formulation is that the
constraints (\ref{6.19}) become conditions for G-analyticity. We
introduce harmonic variables describing the coset
$\mbox{USp}(2N)/[\mbox{U}(1)]^N$:
\begin{equation}\label{6.25}
  u\in \mbox{USp}(2N): \qquad u^I_iu^i_J = \delta^I_J\;,
\ \ u^I_i \Omega^{ij}u^J_j = \Omega^{IJ}\;, \ \  u^I_i=
(u^i_I)^*\;.
\end{equation}
Here the indices $i,j$ belong to the fundamental representation of
$\mbox{USp}(2N)$ and $I,J$ are labels corresponding to the
$[\mbox{U}(1)]^N$ projections. The harmonic derivatives
\begin{equation}\label{6.26}
  D^{IJ} = \Omega^{K(I}u^{J)}_i{\partial\over\partial u^K_i}
\end{equation}
form the algebra of $\mbox{USp}(2N)_R$ (see (\ref{6.4})) realized
on the indices $I,J$.

Let us now project the defining constraint (\ref{6.19}) with the
harmonics $u^K_k u^1_{i_1}\ldots u^n_{i_n}$, $K=1,\ldots,n$:
\begin{equation}\label{6.27}
D^1_\alpha W^{12\ldots n} = D^2_\alpha W^{12\ldots n}= \ldots =
D^n_\alpha W^{12\ldots n} =0
\end{equation}
where $D^{K}_\alpha = D^i_\alpha u^{K}_i$ and $W^{12\ldots
n}=W^{\{i_1\ldots i_n\}}u^1_{i_1}\ldots u^n_{i_n}$. Indeed, the
constraint (\ref{6.19}) now takes the form of a G-analyticity
condition. In the appropriate basis in superspace the solution to
(\ref{6.27}) is a short superfield depending on part of the odd
coordinates:
\begin{equation}\label{6.28}
W^{12\ldots n}(x_A,\theta^1,\theta^2,\ldots, \theta^{2N-n},u)\;.
\end{equation}
In addition to (\ref{6.27}), the projected superfield $W^{12\ldots
n}$ automatically satisfies the $\mbox{USp}(2N)$ harmonic
irreducibility conditions
\begin{equation}\label{6.29}
   D^{K\; 2N-K}W^{12} = 0\;, \quad K=1,\ldots,N
\end{equation}
(only the simple roots of $\mbox{USp}(2N)$ are shown). The
equivalence between the two forms of the constraint follows from
the obvious properties of the harmonic products $u^K_{[k} u^K_{i]}
=0$ and $\Omega^{ij}u^K_iu^L_j=0$ for $1\leq K < L\leq n$. The
harmonic constraints (\ref{6.29}) make the superfield ultrashort.

Finally, in case (ii), projecting the constraint (\ref{6.23}) with
$u^I_iu^I_j$ where $I=1,\ldots,N$ (no summation), we obtain the
condition
\begin{equation}\label{6.30'}
  D^I_\alpha D^I_\beta w=0\;.
\end{equation}
It implies that the superfield $w$ is {\sl linear} in each
projection $\theta^{\alpha I}$.

\subsection{Series of UIR's of $\mbox{OSp}(8^*/2N)$ and shortening}
\label{short6}

It is now clear that we can realize the BPS series of UIR's
(\ref{6.18}) as products of the different G-analytic superfields
(supersingletons) (\ref{6.27}).\footnote{As a bonus, we also
prove the unitarity of these series, since they are obtained by
multiplying massless unitary multiplets.} BPS shortening is
obtained by setting the first $p-1$ $\mbox{USp}(2N)$ Dynkin
labels to zero:
\begin{equation}\label{6.34}
{p\over 2N}\ \mbox{BPS}:\ \ W^{[0,\ldots,0,a_p,\ldots,a_N]}
(\theta^1,\theta^2,\ldots,\theta^{2N-p}) =  (W^{1\ldots
p})^{a_p}\ldots (W^{1\ldots N})^{a_N}\;.
\end{equation}

We remark that our harmonic coset
$\mbox{USp}(2N)/[\mbox{U}(1)]^N$ is effectively reduced to
$\mbox{USp}(2N)/\mbox{U}(p)\times [\mbox{U}(1)]^{N-p}$ in the
case of $p/2N$ BPS shortening. Such a smaller harmonic space was
used in  Ref. \cite{Howe} to formulate the $(2,0)$ tensor
multiplet.

A study of the most general UIR's of $\mbox{OSp}(8^*/2N)$
(similar to the one of Ref. \cite{dp} for the case of
$\mbox{SU}(2,2/N)$) is presented in Ref. \cite{Minw2}. We can
construct these UIR's by multiplying the three types of
supersingletons above:
\begin{equation}\label{6.32}
  w_{\alpha_1\ldots\alpha_{m_1}}w_{\beta_1\ldots\beta_{m_2}}
w_{\gamma_1\ldots\gamma_{m_3}}\; w^k\; W^{[a_1,\ldots,a_N]}
\end{equation}
where $m_1\geq m_2 \geq m_3$ and the spinor indices are arranged
so that they form an $\mbox{SU}^*(4)$ UIR with Young tableau
$(m_1,m_2,m_3)$ or Dynkin labels
$J_1=m_1-m_2,J_2=m_2-m_3,J_3=m_3$. Thus we obtain four distinct
series:
\begin{eqnarray}
  \mbox{A)}&& \ell
\geq 6 +{1\over 2}(J_1+2J_2+3J_3)+2\sum_{k=1}^N a_k\;; \nonumber\\
  \mbox{B)}&& J_3=0\;,  \qquad \ell
\geq 4 +{1\over 2}(J_1+2J_2)+2\sum_{k=1}^N a_k\;; \nonumber\\
  \mbox{C)}&& J_2=J_3=0\;, \qquad \ell
\geq 2 +{1\over 2}J_1+2\sum_{k=1}^N a_k\;; \nonumber\\
  \mbox{D)}&& J_1=J_2=J_3= 0\;, \qquad \ell
= 2\sum_{k=1}^N a_k\;. \label{6.33'}
\end{eqnarray}
The superconformal bound is saturated when $k=0$ in (\ref{6.32}).
Note that the values of the conformal dimension we can obtain are
``quantized" since the factor $w^k$ has $\ell=2k$ and $k$ must be
a non-negative integer to ensure unitarity. With this restriction
eq. (\ref{6.33'}) reproduces the results of Ref. \cite{Minw2}.
However, we cannot comment on the existence of a ``window" of
dimensions $2 +{1\over 2}J_1+2\sum_{k=1}^N a_k\leq \ell \leq 4
+{1\over 2}J_1+2\sum_{k=1}^N a_k$ conjectured in \cite{Minw2}.
\footnote{In a recent paper \cite{FFr} the UIR's of the
six-dimensional conformal algebra $\mbox{SO}(2,6)$ have been
classified. Note that the superconformal bound in case A (with all
$a_i=0$) is stronger that the purely conformal unitarity bounds
found in \cite{FFr}.}

In the generic case the multiplet (\ref{6.32}) is ``long", but for
certain special values of the dimension some shortening can take
place \cite{Minw2}. We can immediately identify all these short
multiplets. First of all, case D corresponds to BPS shortening.
In the other cases let us first set $a_i=0$, i.e. no BPS
multiplets appear in (\ref{6.32}). Then saturating the bound in
case A (i.e., setting $k=0$) leads to the shortening condition
(see (\ref{6.24})):
\begin{equation}\label{6.3311}
  \epsilon^{\delta\alpha\beta\gamma}
D^i_\delta(w_{\alpha\ldots\alpha_{m_1}}w_{\beta\ldots\beta_{m_2}}
w_{\gamma\ldots\gamma_{m_3}}) = 0\ \rightarrow \ \ell = 6
+{1\over 2}(J_1+2J_2+3J_3)\;.
\end{equation}
Next, in case B we have two possibilities: either we saturate the
bound ($k=0$) or we use just one factor $w$ ($k=1$). Using
(\ref{6.23}) and (\ref{6.24}), we find
\begin{equation}\label{6.3312}
  \epsilon^{\delta\gamma\alpha\beta} D^i_\gamma(w_{\alpha\ldots\alpha_{m_1}}
w_{\beta\ldots\beta_{m_2}}) = 0\ \rightarrow \ \ell = 4 +{1\over
2}(J_1+2J_2)\;;
\end{equation}
\begin{equation}\label{6.3313}
 \epsilon^{\delta\gamma\alpha\beta}D^{(i}_\delta
D^{j)}_\gamma(w\;w_{\alpha\ldots\alpha_{m_1}}
w_{\beta\ldots\beta_{m_2}}) = 0\ \rightarrow \ \ell = 6 +{1\over
2}(J_1+2J_2)\;.
\end{equation}
Similarly, in case C with $J_1\neq0$ we have three options, namely
setting $k=0\ \rightarrow \ \ell = 2+{1\over 2}J_1$ (which
corresponds to the supersingleton defining constraint
(\ref{6.24})) or $k=1,2$ which gives:
\begin{equation}\label{6.3314}
 \epsilon^{\delta\gamma\beta\alpha}D^{(i}_\gamma
D^{j)}_\beta(w\;w_{\alpha\ldots\alpha_{m_1}}) = 0\ \rightarrow \
\ell = 4 +{1\over 2}J_1\;,
\end{equation}
\begin{equation}\label{6.3315}
 \epsilon^{\delta\gamma\beta\alpha_1}D^{(i}_\delta D^j_\gamma
D^{k)}_\beta(w^2\;w_{\alpha_1\ldots\alpha_{m_1}}) = 0\ \rightarrow
\ \ell = 6 +{1\over 2}J_1\;.
\end{equation}
Finally, in case C with $J_1=0$ we can take the scalar
supersingleton (\ref{6.23}) itself, i.e. set $k=1\ \rightarrow \
\ell = 2$, or set $k=2,3$:
\begin{equation}\label{6.3316}
  \epsilon^{\delta\gamma\beta\alpha}D^{(i}_\gamma D^j_\beta
D^{k)}_\alpha(w^2) = 0\ \rightarrow \ \ell = 4\;,
\end{equation}
\begin{equation}\label{6.3317}
  \epsilon^{\delta\gamma\beta\alpha}D^{(i}_\delta D^j_\gamma
D^k_\beta D^{l)}_\alpha(w^3) = 0\ \rightarrow \ \ell = 6\;.
\end{equation}

Introducing $\mbox{USp}(2N)$ quantum numbers into the above
shortening conditions is achieved by multiplying the short
multiplets by a BPS object. The new short multiplets satisfy the
corresponding $\mbox{USp}(2N)$ projections of eqs. (\ref{6.23}),
(\ref{6.24}), (\ref{6.3311})-(\ref{6.3317}). We call such objects
``intermediate short".
 \vfill\eject

\setcounter{equation}0
\section{The three-dimensional case}

In this section we carry out the analysis of the $d=3$ $N=8$
superconformal algebra $\mbox{OSp}(8/4,\mathbb{R})$ in a way
similar to the above (the generalization to
$\mbox{OSp}(N/4,\mathbb{R})$ is straightforward). Some of the
short representations of the $N=2$ and $N=3$ cases were discussed
in Ref. \cite{Torino}.

\subsection{The conformal superalgebra $\mbox{OSp}(8/4,\mathbb{R})$
and G-analyticity}\label{CSGA}

The part of the conformal superalgebra
$\mbox{OSp}(8/4,\mathbb{R})$ relevant to our discussion is given
below:
\begin{eqnarray}
  && \{Q^i_\alpha, Q^j_\beta\} = 2\delta^{ij} \gamma^\mu_{\alpha\beta}
P_\mu\;, \label{7.1}\\
  && \{Q^i_\alpha, S^j_\beta\} = \delta^{ij}
M_{\alpha\beta} + 2\epsilon_{\alpha\beta}  (T^{ij} +  \delta^{ij}
D) \;, \label{7.2}\\
  && [T^{ij}, Q^k_\alpha] = i(\delta^{ki} Q^j_\alpha - \delta^{kj}
Q^i_\alpha)\;,\label{7.3}\\
  && [T^{ij}, T^{kl}] = i(\delta^{ik} T^{jl} + \delta^{jl} T^{ik}
- \delta^{jk} T^{il} - \delta^{il} T^{jk})\;. \label{7.4}
\end{eqnarray}
Here we find the generators $Q^i_\alpha$ of $N=8$ Poincar\'{e}
supersymmetry with a spinor index $\alpha=1,2$ of the $d=3$
Lorentz group $\mbox{SL}(2,\mathbb{R})\sim \mbox{SO}(1,2)$
(generators $M_{\alpha\beta} = M_{\beta\alpha}$) and a
vector\footnote{Ascribing one of the three 8-dimensional
representations of $\mbox{SO}(8)$, $8_v$, $8_s$, $8_c$ (related
by triality) to the supersymmetry generators is purely
conventional. Since in all the other $N$-extended $d=3$
supersymmetries the odd generators belong to the vector
representation, we prefer to put an $8_v$ index $i$ on the
supercharges.} index $i=1,\ldots,8$ of the R symmetry group
$\mbox{SO}(8)$ (generators $T^{ij}=-T^{ji}$); $S^i_\alpha$ of
conformal supersymmetry; $P_\mu$, $\mu=0,1,2$, of translations;
$D$ of dilations.

The standard realization of this superalgebra makes use of a
superspace with coordinates $x^\mu,\theta^{\alpha\; i}$. In order
to study G-analyticity we need to decompose the generators
$Q^i_\alpha$ under $[\mbox{U}(1)]^4\subset \mbox{SO}(8)$. Besides
the vector representation $8_v$ of $\mbox{SO}(8)$ we are also
going to use the spinor ones, $8_s$ and $8_c$. Denoting the four
$\mbox{U}(1)$ charges by $\pm$, $(\pm)$, $[\pm]$ and $\{\pm\}$,
we decompose the three 8-dimensional representations as follows:
\begin{eqnarray}
8_v:\quad Q^i &\rightarrow& Q^{\pm\pm}, \ Q^{(\pm\pm)}, \
Q^{[\pm]\{\pm\}},\label{7.10}\\
 8_s:\quad \phi^a &\rightarrow&
\phi^{+(+)[\pm]}, \ \phi^{-(-)[\pm]}, \ \phi^{+(-)\{\pm\}}, \
\phi^{-(+)\{\pm\}}\label{7.11}\\
 8_c:\quad  \sigma^{\dot a} &\rightarrow& \sigma^{+(+)\{\pm\}},
\ \sigma^{-(-)\{\pm\}}, \ \sigma^{+(-)[\pm]}, \
\sigma^{-(+)[\pm]}\label{7.12}
\end{eqnarray}
The definition of the charge operators $H_i$, $i=1,2,3,4$ can be
read off from the corresponding projections of the relation
(\ref{7.2}):
\begin{eqnarray}
 \{Q^{++}_\alpha, S^{--}_\beta\} &=& {1\over 2} M_{\alpha\beta} +
\epsilon_{\alpha\beta}  (D-{1\over 2}H_1) \;, \nonumber\\
  \{Q^{(++)}_\alpha, S^{(--)}_\beta\} &=& {1\over 2} M_{\alpha\beta} +
\epsilon_{\alpha\beta}  (D - {1\over 2}H_2) \;, \nonumber\\
   \{Q^{[+]\{+\}}_\alpha, S^{[-]\{-\}}_\beta\} &=& {1\over 2} M_{\alpha\beta} +
\epsilon_{\alpha\beta}  (D - {1\over 2}H_3  - {1\over 2}H_4) \;,
\nonumber\\
   \{Q^{[+]\{-\}}_\alpha, S^{[-]\{+\}}_\beta\} &=& -{1\over 2} M_{\alpha\beta} -
\epsilon_{\alpha\beta}  (D - {1\over 2}H_3  + {1\over 2}H_4) \; .
\label{7.9}
\end{eqnarray}

Let us denote a quasi primary superconformal field of the
$\mbox{OSp}(8/4,\mathbb{R})$ algebra by the quantum numbers of its
HWS, ${\cal D}(\ell; J; a_1,a_2,a_3,a_4)$, where $\ell$ is the
conformal dimension, $J$ is the Lorentz spin and $a_i$ are the
Dynkin labels (see, e.g., \cite{FSS}) of the $\mbox{SO}(8)$ R
symmetry. In fact, in our scheme the natural labels are the four
charges $h_i$ (the eigenvalues of $H_i$) which are related to the
Dynkin labels as follows: $a_1= {1\over 2}(h_1-h_2)\;, \quad a_2=
{1\over 2}(h_2-h_3-h_4)\;, \quad a_3=h_3\;, \quad a_4=h_4$. A HWS
$|a_i\rangle$ of $\mbox{SO}(8)$ is by definition annihilated by
the positive simple roots of the $\mbox{SO}(8)$ algebra:
\begin{equation}\label{7.15}
  T^{[++]}|a_i\rangle = T^{\{++\}}|a_i\rangle =
T^{++(--)}|a_i\rangle = T^{(++)[-]\{-\}}|a_i\rangle = 0\;.
\end{equation}

G-analyticity is obtained by requiring that one or more
projections of $Q^i_\alpha$ annihilate the state. These
projections must form an anticommuting subset closed under the
action of the raising operators of $\mbox{SO}(8)$ (\ref{7.15}).
Then, using the algebra (\ref{7.9}) we examine the consistency of
the G-analyticity conditions with the definition of a
superconformal primary $S^i_\alpha|\ell; J; a_k\rangle=0$. Thus we
find the following set of G-analytic superspaces corresponding to
BPS states:
\begin{eqnarray}
 {1\over 8}\mbox{ BPS:} && \left\{
  \begin{array}{l}
    q_\alpha^{++}\Phi=0\ \rightarrow \
    \Phi(\theta^{++},\theta^{(\pm\pm)},\theta^{[\pm]\{\pm\}})\\
    {\cal D}(a_1+a_2 + {1\over 2}(a_3+a_4);0;a_1,a_2,a_3,a_4)
     \end{array}
 \right.\nonumber\\
 {1\over 4}\mbox{ BPS:}  &&\left\{
  \begin{array}{l}
    q_\alpha^{++}\Phi=q_\alpha^{(++)}\Phi=0\ \rightarrow \
    \Phi(\theta^{++},\theta^{(++)},\theta^{[\pm]\{\pm\}})\\
    {\cal D}(a_2 + {1\over 2}(a_3+a_4);0;0,a_2,a_3,a_4)
      \end{array}
 \right. \label{7.20}\\
{3\over 8}\mbox{ BPS:}   &&\left\{
  \begin{array}{l}
    q_\alpha^{++}\Phi=q_\alpha^{(++)}\Phi=q_\alpha^{[+]\{+\}}\Phi=0\ \rightarrow
    \   \Phi(\theta^{++},\theta^{(++)},\theta^{[+]\{\pm\}},\theta^{[-]\{+\}})\\
    {\cal D}({1\over 2}(a_3+a_4);0;0,0,a_3,a_4)
  \end{array}
 \right. \nonumber\\
{1\over 2}  \mbox{ BPS (I):} &&\left\{
  \begin{array}{l}
    q_\alpha^{++}\Phi=q_\alpha^{(++)}\Phi=q_\alpha^{[+]\{\pm\}}\Phi=0\ \rightarrow \
    \Phi(\theta^{++},\theta^{(++)},\theta^{[+]\{\pm\}})\\
    {\cal D}({1\over 2}a_3;0;0,0,a_3,0)
  \end{array}
 \right. \nonumber\\
{1\over 2}  \mbox{ BPS (II):} &&\left\{
  \begin{array}{l}
    q_\alpha^{++}\Phi=q_\alpha^{(++)}\Phi=q_\alpha^{[\pm]\{+\}}\Phi=0\ \rightarrow \
    \Phi(\theta^{++},\theta^{(++)},\theta^{[\pm]\{+\}})\\
    {\cal D}({1\over 2}a_4;0;0,0,0,a_4)
  \end{array}
 \right. \nonumber
\end{eqnarray}

We remark that the states $1/4$, $3/8$ and $1/2$ are annihilated
by some of the lowering operators of $\mbox{SO}(8)$. This means
that the $\mbox{SO}(8)$ subalgebras $\mbox{SU}(2)$,
$\mbox{SU}(3)$ and $\mbox{SU}(4)$, respectively, act trivially.
These properties are equivalent to the restrictions on the
possible values of the $\mbox{SO}(8)$ Dynkin labels in
(\ref{7.20}). Note the existence of two types of $1/2$ BPS states
due to the two possible subsets of projections of $q^i$ closed
under the raising operators of $\mbox{SO}(8)$ (\ref{7.15}). This
fact can be equivalently explained by the two possible embeddings
of $\mbox{SU}(4)$ in $\mbox{SO}(8)$.

\subsection{Supersingletons and harmonic superspace}

The supersingletons are the simplest $\mbox{OSp}(8/4,\mathbb{R})$
representations of the $1/2$ type in (\ref{7.20}) and correspond
to ${\cal D}(1/2; 0; 0,0,1,0)$ or ${\cal D}(1/2; 0; 0,0,0,1)$.
The existence of two distinct types of $d=3$ $N=8$
supersingletons has first been noted in Ref. \cite{GNST}. Each of
them is just a collection of eight Dirac supermultiplets \cite{Fr}
made out of ``Di" and ``Rac" singletons \cite{ff2}.

In order to realize the supersingletons in superspace we note that
the HWS in the two supermultiplets above has spin 0 and the Dynkin
labels of the $8_s$ or $8_c$ of $\mbox{SO}(8)$, correspondingly.
Thus, we take a scalar superfield $\Phi_a(x^\mu,
\theta^\alpha_i)$ (or $\Sigma_{\dot a}(x^\mu, \theta^\alpha_i)$)
with an external $8_s$ index $a$ (or an $8_c$ index $\dot a$).
These superfields are subject to the following on-shell
constraints \footnote{See also \cite{Howe} for the description of
a supersingleton related to ours by $\mbox{SO}(8)$ triality.
Superfield representations of other $OSp(N/4)$ superalgebras were
considered in \cite{IS,FFre}.}:
\begin{eqnarray}
  \mbox{type I:}&&D^i_\alpha\Phi_a = {1\over 8}\gamma^i_{a\dot
b}\tilde\gamma^j_{\dot b c} D^j_\alpha\Phi_c\;; \label{7.25}\\
  \mbox{type II:}&& D^i_\alpha\Sigma_{\dot a} = {1\over
8}\tilde\gamma^i_{\dot a b}\gamma^j_{b\dot c}
D^j_\alpha\Sigma_{\dot c} \label{7.26}
\end{eqnarray}
which reduce them to a massless $8_s$ ($8_c$) scalar  and $8_c$
($8_s$)  spinor.

The harmonic superspace description of these supersingletons can
be realized by taking the harmonic coset\footnote{A formulation
of the above multiplet in harmonic superspace has been proposed in
Ref. \cite{Howe} (see also \cite{ZK} and \cite{HL} for a general
discussion of three-dimensional harmonic superspaces). The
harmonic coset used in \cite{Howe} is
$\mbox{Spin}(8)/\mbox{U}(4)$. Although the supersingleton itself
does indeed live in this smaller coset, its residual symmetry
$U(4)$ would not allow us to multiply different realizations of
the supersingleton. For this reason we prefer from the very
beginning to use the coset ${\mbox{Spin}(8)/[\mbox{U}(1)]^4}$
with a minimal residual symmetry.} $ {\mbox{SO}(8)/
[\mbox{SO}(2)]^4} \ \sim \ {\mbox{Spin}(8)/[\mbox{U}(1)]^4}$.
Since $\mbox{SO}(8)$  has three inequivalent fundamental
representations, $8_s,8_c,8_v$, following  \cite{GHS} we
introduce three sets of harmonic variables, $u_a^A\;, \ w^{\dot
A}_{\dot a}\;, \ v^I_i$, where $A$, $\dot A$ and $I$ denote the
decompositions of an $8_s$, $8_c$ and $8_v$ index,
correspondingly, into sets of four $\mbox{U}(1)$ charges (see
(\ref{7.10})-(\ref{7.12})). Each of these $8\times 8$ real
matrices belongs to the corresponding representation of
$\mbox{SO}(8)$. This implies that they are orthogonal matrices:
\begin{equation}\label{7.28}
  u_a^A u_a^B = \delta^{AB}\;, \quad w^{\dot A}_{\dot a} w^{\dot B}_{\dot
a}  = \delta^{\dot A\dot B}\;, \quad v^I_i v^J_i = \delta^{IJ} \;.
\end{equation}
These matrices supply three copies of the group space, and we only
need one to parametrize the harmonic coset. The condition which
identifies the three sets of harmonic variables is
\begin{equation}\label{7.29}
  u_a^A (\gamma^I)_{A\dot A} w^{\dot A}_{\dot a} = v^I_i (\gamma^i)_{a\dot
a}\;.
\end{equation}

Further, we introduce harmonic derivatives (the covariant
derivatives on the coset ${\mbox{Spin}(8)/[\mbox{U}(1)]^4}$):
\begin{equation}\label{7.30}
  D^{IJ} = u^A_a (\gamma^{IJ})^{AB}{\partial\over\partial u^B_a} +
w^{\dot A}_{\dot a} (\gamma^{IJ})^{\dot A\dot
B}{\partial\over\partial w^{\dot B}_{\dot a}} + v^{[I}_i
{\partial\over\partial v^{J]}_{i}}\;.
\end{equation}
They respect the algebraic relations  (\ref{7.28}), (\ref{7.29})
among the harmonic variables and form the algebra of
$\mbox{SO}(8)$ realized on the indices $A,\dot A, I$.

We now use the harmonic variables for projecting the
supersingleton defining constraints (\ref{7.25}), (\ref{7.26}).
From (\ref{7.29}) it follows that the projections $\Phi^{+(+)[+]}$
and $\Sigma^{+(+)\{+\}}$ satisfy the following G-analyticity
constraints:
\begin{eqnarray}
  &&D^{++}\Phi^{+(+)[+]} = D^{(++)}\Phi^{+(+)[+]}=D^{[+]\{\pm\}}
\Phi^{+(+)[+]} = 0\;, \label{7.31}\\
  &&D^{++}\Sigma^{+(+)\{+\}} = D^{(++)}\Sigma^{+(+)\{+\}}=D^{[+]\{\pm\}}
\Sigma^{+(+)\{+\}} = 0 \label{7.31'}
\end{eqnarray}
where $D^I_\alpha = v^I_iD^i_\alpha$, $\Phi^A = u^A_a\Phi_a$ and
$\Sigma^{\dot A} = w^{\dot A}_{\dot a}\Sigma_{\dot a}$. This is
the superspace realization of the 1/2 BPS shortening conditions in
(\ref{7.20}). In the appropriate basis in superspace
$\Phi^{+(+)[+]}$ and $\Sigma^{+(+)\{+\}}$ depend on different
halves of the odd variables as well as on the harmonic variables:
\begin{eqnarray}
  \mbox{type I}:&& \Phi^{+(+)[+]}
(x_A,\theta^{++}, \theta^{(++)}, \theta^{[+]\{\pm\}}, u,w) \;,
\label{7.32}\\
 \mbox{type II}: && \Sigma^{+(+)\{+\}}(x_A,\theta^{++},\theta^{(++)},
\theta^{[\pm]\{+\}}, u,w)\;.\label{7.32'}
\end{eqnarray}

In addition to the G-analyticity constraints (\ref{7.31}),
(\ref{7.31'}), the on-shell superfields $\Phi^{+(+)[+]}$,
$\Sigma^{+(+)\{+\}}$ are subject to the $\mbox{SO}(8)$
irreducibility harmonic conditions obtained from (\ref{7.15}) by
replacing the $\mbox{SO}(8)$ generators by the corresponding
harmonic derivatives. The combination of the latter with eq.
(\ref{7.31}) is equivalent to the original constraint
(\ref{7.25}).

Note that $\Phi^{+(+)[+]}$, $\Sigma^{+(+)\{+\}}$ are automatically
annihilated by some of the lowering operators of $\mbox{SO}(8)$.
This means that the supersingleton harmonic superfields
effectively live on the smaller harmonic coset
$\mbox{Spin}(8)/\mbox{U}(4)$.

\subsection{$\mbox{OSp}(8/4,\mathbb{R})$ supersingleton composites}

One way to obtain short multiplets of $\mbox{OSp}(8/4,\mathbb{R})$
is to multiply different analytic superfields describing the type
I supersingleton. The point is that above we chose a particular
projection of, e.g., the defining constraint (\ref{7.25}) which
lead to the analytic superfield  $\Phi^{+(+)[+]}$. In fact, we
could have done this in a variety of ways, each time obtaining
superfields depending on different halves of the total number of
odd variables. Thus, we can have four distinct but equivalent
analytic descriptions of the type I supersingleton:
\begin{eqnarray}
  &&\Phi^{+(+)[+]}
(\theta^{++}, \theta^{(++)}, \theta^{[+]\{+\}},
\theta^{[+]\{-\}})\;, \nonumber\\
  &&\Phi^{+(+)[-]} (\theta^{++}, \theta^{(++)},
\theta^{[-]\{+\}}, \theta^{[-]\{-\}})\;, \nonumber\\
  &&\Phi^{+(-)\{+\}} (\theta^{++},
\theta^{(--)}, \theta^{[+]\{+\}}, \theta^{[-]\{+\}})\;,
\nonumber\\
  &&\Phi^{+(-)\{-\}}
(\theta^{++}, \theta^{(--)}, \theta^{[+]\{-\}},
\theta^{[-]\{-\}})\;. \label{7.34}
\end{eqnarray}
Then we can multiply them in the following way:
\begin{equation}\label{7.35}
  (\Phi^{+(+)[+]})^{p+q+r+s}(\Phi^{+(+)[-]})^{q+r+s}
(\Phi^{+(-)\{+\}})^{r+s}(\Phi^{+(-)\{-\}})^{s}
\end{equation}
thus obtaining three BPS series of $\mbox{OSp}(8/4,\mathbb{R})$
UIR's:
\begin{eqnarray}
 {1\over 8}  \mbox{ BPS:} && {\cal D}(a_1+a_2 + {1\over
2}(a_3+a_4), 0; a_1,a_2,a_3,a_4)\;, \quad a_1-a_4 = 2s \geq 0\;;
 \nonumber\\
 {1\over 4}  \mbox{ BPS:} && {\cal D}(a_2 + {1\over 2}a_3, 0;
0,a_2,a_3,0)\;;  \label{7.36}\\
 {1\over 2}  \mbox{ BPS:} && {\cal D}({1\over 2}a_3, 0; 0,0,a_3,0)
\nonumber \end{eqnarray} where $a_1=r+2s\;, \  a_2= q\;, \
a_3=p\;, \  a_4=r\;$.

We see that using only one type of supersingletons cannot
reproduce the classification (\ref{7.20}), in particular, the
$3/8$ series. The latter can be obtained  by mixing the two types
of supersingletons:
\begin{equation}\label{7.37}
[\Phi^{+(+)[+]}(\theta^{++},\theta^{(++)},\theta^{[+]\{\pm\}})]^{a_3}
[\Sigma^{+(+)\{+\}}(\theta^{++},\theta^{(++)},\theta^{[\pm]\{+\}})]^{a_4}
\end{equation}
(or the same with $\Phi$ and $\Sigma$ exchanged). Counting the
charges and the dimension, we find exact matching with the $3/8$
series in (\ref{7.20}). Further, mixing two realizations of type I
and one of type II supersingletons, we can construct the 1/4
series in (\ref{7.20}):
\begin{equation}\label{7.38}
   [\Phi^{+(+)[+]}]^{a_2+a_3}[\Phi^{+(+)[-]}]^{a_2}
[\Sigma^{+(+)\{+\}}]^{a_4}\;.
\end{equation}
Finally, the full 1/8 series in (\ref{7.20}) (i.e., without the
restriction $a_1-a_4 = 2s\geq 0$ in (\ref{7.36})) can be obtained
in a variety of ways.

In this section we have analyzed all short highest-weight UIR's of
the $\mbox{OSp}(8/4,\mathbb{R})$ superalgebra whose HWS's are
annihilated by part of the super-Poincar\'{e} odd generators. The
number of distinct possibilities have been shown to correspond to
different BPS conditions on the HWS. When the algebra is
interpreted on the $AdS_4$ bulk, for which the 3d superconformal
field theory corresponds to the boundary M-2 brane dynamics,
these states appear as BPS massive excitations, such as K-K
states or AdS black holes, of M-theory on $AdS_4\times S^7$.
Since in M-theory there is only one type of supersingleton
related to the M-2 brane transverse coordinates \cite{Duff1},
according to our analysis massive states cannot be 3/8 BPS
saturated, exactly as it happens in M-theory on $M^4\times T^7$.
Indeed, the missing solution was also noticed in Ref.
\cite{Duff2} by studying $AdS_4$ black holes in gauged $N=8$
supergravity. Curiously, in the ungauged theory, which is in some
sense the flat limit of the former, the 3/8 BPS states are
forbidden \cite{FMG} by the underlying $E_{7(7)}$ symmetry of
$N=8$ supergravity \cite{CJ}.

\subsection{Series of UIR's of $\mbox{OSp}(8/4,\mathbb{R})$}\label{short3}

In the even-dimensional case $d=6$ we had supersingleton
superfields carrying either $R$ symmetry indices or Lorentz
indices or just conformal dimension. Multiplying them we were
able to reproduce the corresponding general series of UIR's. In
the odd-dimensional case $d=3$ we only have two supersingletons
carrying $\mbox{SO}(8)$ spinor indices. Multiplying them we could
construct all the short objects of BPS type. Yet, for reproducing
the most general UIR's (see \cite{Minw2}), we need short objects
with spin but without $\mbox{SO}(8)$ indices. These arise in the
form of conserved currents. The simplest one is a Lorentz scalar
and an $\mbox{SO}(8)$ singlet $w$ of dimension $\ell=1$. It can
be realized as a bilinear of two supersingletons of the same
type, e.g., $w=\Phi_a \Phi_a$ or $w=\Sigma_{\dot a}\Sigma_{\dot
a}$. Using (\ref{7.25}) or (\ref{7.26}) one can show that it
satisfies the constraint (a non-BPS shortness condition)
\begin{equation}\label{7.001}
  D^i_\alpha D^{j\alpha} \; w =
{1\over 8}\delta^{ij} D^k_\alpha D^{k\alpha} \; w \;.
\end{equation}
The other currents carry $\mbox{SL}(2,\mathbb{R})$ spinor indices,
$w_{\alpha_1\ldots\alpha_{2J}}$, have dimension $\ell=1 + J$ and
satisfy the constraint \cite{Park3d}
\begin{equation}\label{7.002}
  D^{i\alpha}w_{\alpha\alpha_2\ldots\alpha_{2J}} = 0\;.
\end{equation}
They can be constructed as bilinears of the two types of
supersingletons (for half-integer spin) or of  two copies of the
same type (for integer spin). For example, the two lowest ones
($J=1/2$ and $J=1$) are
\begin{equation}\label{7.0020}
  w_\alpha = \gamma^i_{b\dot b}\left(D^i_\alpha \Phi_b\Sigma_{\dot b} -
\Phi_b D^i_\alpha \Sigma_{\dot b} \right)\;,
\end{equation}
\begin{equation}\label{7.0021}
  w_{\alpha\beta} =  D^i_{(\alpha}\Phi_a  (\gamma^i\gamma^j)_{ab}
D^j_{\beta)}\Phi'_b + 32i(\Phi_a\partial_{\alpha\beta}\Phi'_a -
\partial_{\alpha\beta}\Phi_a \Phi'_a)\;.
\end{equation}

The generic ``long" UIR of $\mbox{OSp}(8/4,\mathbb{R})$ can now be
obtained as a product of all of the above short objects:
\begin{equation}\label{7.003}
  w_{\alpha_1\ldots\alpha_{2J}}\; w^k\;
\mbox{BPS}[a_1,a_2,a_3,a_4]\;.
\end{equation}
Here we have used the first factor to obtain the spin, the second
one for the conformal dimension and the BPS factor for the
$\mbox{SO}(8)$ quantum numbers. The unitarity bound is given by
\begin{equation}\label{7.004}
  \ell \geq 1+J+a_1+a_2+{1\over 2}(a_3+a_4)
\end{equation}
and is saturated if $k=0$ in (\ref{7.003}). The object
(\ref{7.003}) is short if: (i) $J\neq 0$ and $k=0$ (then it
satisfies the intersection of (\ref{7.002}) with the BPS
conditions); (ii) $J=0$ and $k=1$  (then it satisfies the
intersection of (\ref{7.001}) with the BPS conditions); (iii)
$J=0$ and $k=0$ (then it is BPS short). These results exactly
match the classification of Ref. \cite{Minw2}.

\setcounter{equation}0
\section{Conclusions}

Here we give a summary of the different types of BPS states which
are realized as products of supersingletons described by
G-analytic harmonic superfields. We shall restrict ourselves to
the physically interesting cases of $M_2$ and $M_5$ branes
horizon geometry where only one type of such supersingletons
appears. This construction gives rise to a restricted class of the
most general BPS states.

\subsection{$\mbox{OSp}(8^*/4)$}

The BPS states are constructed in terms of the $(2,0)$ $d=6$
tensor multiplet $W^{\{ij\}}$ in two equivalent G-analytic
realizations:
\begin{equation}\label{8.02}
  (W^{12}(\theta^{1,2})^{p+q}
(W^{13}(\theta^{1,3}))^{q} \;.
\end{equation}

\begin{table}[h]
  \begin{center}
    \leavevmode
\label{bps6}
    \begin{tabular}{llll}
 BPS & USp(4) & Dimension & Harmonic space \\ \hline
  \\
 ${1\over 2}$ & (0,p) & 2p & USp(4)/U(2) \\
  \\
 ${1\over 4}$ & (2q,p) & 2p+4q & USp(4)/[U(1)$]^2$ \\
              & (2q,0) & 4q & USp(4)/U(2) \\
    \end{tabular}
  \end{center}
\end{table}

\subsection{$\mbox{OSp}(8/4,\mathbb{R})$}

The type I BPS states are constructed in terms of the $N=8$ $d=3$
matter multiplet $\Phi_a$ carrying an external $8_s$ $SO(8)$
spinor index in four equivalent G-analytic realizations:
\begin{eqnarray}
  &&[\Phi^{+(+)[+]}
(\theta^{++,(++),[+]\{\pm\}})]^{p+q+r+s}\times \nonumber\\
  &&[\Phi^{+(+)[-]} (\theta^{++,(++),[-]\{\pm\}})]^{q+r+s}\times  \nonumber\\
  &&[\Phi^{+(-)\{+\}} (\theta^{++,(--),[\pm]\{+\}})]^{r+s}\times
\nonumber\\
  &&[\Phi^{+(-)\{-\}}
(\theta^{++,(--),[\pm]\{-\}})]^{s}\;. \label{8.03}
\end{eqnarray}

\begin{table}[h]
  \begin{center}
    \leavevmode
\label{bps3}
    \begin{tabular}{llll}
 BPS & SO(8)  & Dimension & Harmonic space \\ \hline
  \\
 ${1\over 2}$ & (0,0,p,0) & ${1\over 2}p$ & Spin(8)/U(4) \\
  \\
 ${1\over 4}$ & (0,q,p,0) & ${1\over 2}(p+2q)$ & Spin(8)/U(2)$\times$U(2) \\
  \\
 ${1\over 8}$ & (r+2s,q,p,r) & ${1\over 2}(p+2q+3r+4s)$ &
Spin(8)/[U(1)$]^4$ \\
    \end{tabular}
  \end{center}
\end{table}

The type II BPS states are constructed in terms of the $N=8$ $d=3$
matter multiplet $\Sigma_{\dot a}$ carrying an external $8_c$
$SO(8)$ spinor index in four equivalent G-analytic realizations:
\begin{eqnarray}
  &&[\Sigma^{+(+)\{+\}}
(\theta^{++,(++),[\pm]\{+\}})]^{p+q+r+s}\times \nonumber\\
  &&[\Sigma^{+(+)\{-\}} (\theta^{++,(++),[\pm]\{-\}})]^{q+r+s}\times  \nonumber\\
  &&[\Sigma^{+(-)[+]} (\theta^{++,(--),[+]\{\pm\}})]^{r+s}\times
\nonumber\\
  &&[\Sigma^{+(-)[-]}
(\theta^{++,(--),[-]\{\pm\}})]^{s}\;. \label{8.033}
\end{eqnarray}

\begin{table}[h]
  \begin{center}
    \leavevmode
\label{bps33}
    \begin{tabular}{llll}
 BPS & SO(8)  & Dimension & Harmonic space \\ \hline
  \\
 ${1\over 2}$ & (0,0,0,p) & ${1\over 2}p$ & Spin(8)/U(4) \\
  \\
 ${1\over 4}$ & (0,q,0,p) & ${1\over 2}(p+2q)$ & Spin(8)/U(2)$\times$U(2) \\
  \\
 ${1\over 8}$ & (r+2s,q,r,p) & ${1\over 2}(p+2q+3r+4s)$ &
Spin(8)/[U(1)$]^4$ \\
    \end{tabular}
  \end{center}
\end{table}


\section*{Acknowledgements}

E.S. is grateful to the TH Division of
CERN for its kind hospitality. The work of S.F. has been
supported in part by the European Commission TMR programme
ERBFMRX-CT96-0045 (Laboratori Nazionali di Frascati, INFN) and by
DOE grant DE-FG03-91ER40662, Task C.

\end{document}